# Building Multi-Platform User Interfaces with UIML

*Mir Farooq Ali, Manuel A. Pérez-Quiñones, Eric Shell*
660 McBryde Hall, Computer Science Dept.
Virginia Tech
Blacksburg, VA 24061, USA
*{mfali | perez | ershell}@cs.vt.edu*

*Marc Abrams*
Harmonia, Inc.
PO Box 11282
Blacksburg, VA 24062, USA
marc@harmonia.com

**Abstract** There has been a widespread emergence of computing devices in the past few years that go beyond the capabilities of traditional desktop computers. However, users want to use the same kinds of applications and access the same data and information on these appliances that they can access on their desktop computers. The user interfaces for these platforms go beyond the traditional interaction metaphors. It is a challenge to build User Interfaces (UIs) for these devices of differing capabilities that allow the end users to perform the same kinds of tasks. The User Interface Markup Language (UIML) is an XML-based language that allows the canonical description of UIs for different platforms. We describe the language features of UIML that facilitate the development of multi-platform UIs. We also describe the key aspects of our approach that makes UIML succeed where previous approaches failed, namely the division in the representation of a UI, the use of a generic vocabulary, and an integrated development environment specifically designed for transformation-based UI development. Finally we describe the initial details of a multi-step usability engineering process for building multi-platform UI using UIML.

## 1.  INTRODUCTION

It is difficult to develop an application for multi-platform deployment without duplicating development effort. Advances in electronics, communications, and the fast growth of the Internet have made the use of a wide variety of computing devices an everyday occurance. Furthermore, users expect to remotely access their data from any of these devices (e.g. desktop computer, laptop, handheld, phone, etc.). Developers now face the daunting task to build user interfaces that must work in multiple devices. To address the need for a uniform language for building multi-platform applications, we have developed the User Interface Markup Language (UIML). In this paper we discuss our research in creating such language, some of the support tools available, and describe our approach towards creating a new user interface design methodology to build multi-platform user interfaces.

UIML is a single language for building user interfaces for any device. UIML emphasizes the separation of concerns of an interactive application in such a way that moving one program from one platform to another (see our definition of platform in the next section) might require only small or no changes at all. Furthermore, because it is based on XML, it is easy to write transformations that take the language from one abstract representation to a more concrete representation. The tools built around UIML extend the language with the use of transformations that afford the developer the creation of user interfaces with a single language that will execute in multiple platforms.

The approach taken with UIML is that of building an application using a generic vocabulary that could then be *rendered* for multiple platforms. Using a generic vocabulary for desktop applications, for example, the developer can write a program in UIML only once and have it be rendered for Java or HTML. And, within HTML, it can be rendered to different versions of HTML and/or browsers. We present some details of the generic vocabulary and the advantages of this approach in this paper.

Based on our experience with UIML, we have learned that the generic vocabulary is not sufficient to build interfaces for widely varying platforms, such as VoiceXML, handhelds (e.g. Palm PDAs) and desktops. In this paper, we describe the preliminary results of a multi-step process of building multi-platform user interfaces. It involves programming in UIML at different levels of abstractions with the support of transformations that move the developer's work from one representation to the next. This paper presents our initial results on this area.

The language design for UIML was done in 1997 by a group of individuals that went on to start the company Harmonia. At the time, the language was deemed too early for commercial adoption, thus the company development plans were put in hiatus for a few years. Meanwhile research in UIML began at the Computer Science Department at Virginia Tech. At the end of 1999, with the expansion of the Internet and the adoption of HTML and other markup languages, it was deemed appropriate to commercialize UIML. Since then, Harmonia has been creating commercial tools to support the deployment of UIML to commercial environments. In parallel, research continues on UIML at Virginia Tech. Support for UIML continues to grow beyond these two organizations. For example there was a conference devoted to UIML in March 2001 held in France. More information on UIML can be found at http://www.uiml.org. Many of the tools described in the paper are available from

Harmonia at http://www.harmonia.com/.

## 2. TERMINOLOGY

Some terminology that is used in the paper is defined next. An *application* is defined to be the back end logic behind a UI that implements the interaction supported by the user interface. A *device* is a physical object with which an end-user interacts using a user interface, such as a Personal Computer (PC), a hand-held computer (like Palm), a cell-phone, an ordinary desktop telephone or a pager. A *toolkit* is the library or markup language used by the application program to build its UI. A toolkit typically describes the widgets like menus, buttons and scrolling bars including their behavior. In the context of this paper, toolkit implies both markup languages and more traditional APIs for imperative or object-oriented languages. A *Vocabulary* is the set of names, properties and associated behavior for UI elements available in any toolkit

A *platform* is a combination of device, operating system and toolkit. An example of a platform is a PC running Windows 2000 on which applications use the Java Swing toolkit. This definition has to be expanded in the case of HTML to include the version of HTML and the particular web browser being used. For example, Internet Explorer 5.5 running on a PC with Windows NT would be a different platform than Netscape Communicator 6.0 running on the same PC with the same OS since they implement different features of HTML differently. A related term that we will use in this paper is *family*, which indicates a group of platforms that have similar layout features. A *Generic Vocabulary* is the vocabulary shared by all platforms of a *family*

*Rendering* is the process of converting a UIML document into a form that can be presented (e.g. through sight or sound) to an end-user, and with which the user can interact. Rendering can be accomplished in two forms: by compiling UIML into another language (e.g. WML or VoiceXML), or by interpreting UIML, meaning that a program reads UIML and makes call to an API that displays the user interface and allows interaction.

## 3. RELATED WORK

### 3.1 Proliferation of Devices

There has been a widespread proliferation of computing devices in the past few years. These computing devices have different interaction styles, input/output techniques, modalities, characteristics, and contexts of use. Dan Olsen [1], in the broader context of problems existing and emerging due to changes in the basics of interaction techniques, talks about problems in interaction emerging due to the diversity of interactive platforms. He uses the term "chaos" to describe the overall problems. He writes that computers of the future could well be wall-sized, desk-sized, palm-sized or even ear-sized. An immediate outcome of these devices of varying capabilities is the corresponding variety of interaction styles. He mentions that while it would be reasonable to expect these varying devices to have different interaction styles, it would not be acceptable to assume that they would operate independently of each other.

Mobile devices introduce an additional complexity. Peter Johnson [12] outlines four concerns regarding the HCI of mobile systems, one of which is accommodating the diversity and integration of devices, network services and applications. Stephen Brewster, *et al.* [6] talk about the problems associated with the small screen size of hand-held devices. In comparison to desktop computers, hand-held devices will always suffer from a lack of screen real estate, so new metaphors of interaction have to be devised for such devices. Some of the other problems that they mention while dealing with small devices were with navigation and presenting information. Two of the problems we encountered in our own work [3] in building UIs for different platforms were the different layout features and screen sizes associated with each platform and device.

### 3.2 Multi-Platform Development

The area of multi-platform UI development falls under the umbrella of what is being termed as the "variety challenges" [1]. There are new challenges for application and solution developers due to the emergence of a variety of users, a variety of devices and channels, and a variety of roles and functions. We would categorize the problem of multi-platform UI development arising due to the emergence of a variety of devices and channels. This research area is relatively new and there has not been a lot of published literature in this area. There have been some approaches towards solving this problem. Building "plastic interfaces" [7, 23] is one such method in which the UIs are designed to "withstand variations of context of use while preserving usability".

Transcoding [4, 10, 11] is a technique used in the World Wide Web for adaptively converting web-content for the increasingly diverse kinds of devices that are being used these days to access web pages. This process also requires some kind of transformation that occurs on the HTML web page to convert it to the desired format. Transcoding assumes multiple forms. In the simplest form, semantic meaning is inferred from the structure of the web page and the page is transformed using this semantic information. A more sophisticated version of transcoding associates annotations with the structural elements of the web page and the transformation occurs based on these annotations. Another version infers semantics based on a group of web pages.

Although these approaches work, they are not too extensible since it is not always possible to infer semantic information from the structural elements of web pages.

## 3.3 Model-Based tools

It is useful to revisit some of the concepts behind model-based UI development tools since we feel that some of these concepts have to be utilized for generating multi-platform UIs. Model-based user interface development tools use different kinds of high-level specification of the tasks that users need to perform, data models that capture the structure and relationships of the information that applications manipulate, specifications of the presentation and dialogue, user models etc, and automatically generate some parts or the complete user interface [14]. One of the central ideas behind model-based tools is to achieve balance between the detailed control of the design of the UI and automation. Many model-based tools have been built in the late 80s and early 90s that include UIDE [19, 20], Interactive UIDE [8], HUMANOID [13, 21], MASTERMIND [22], ITS [25], Mecano [17], and Mobi-D [18].

The central component in a model-based tool is the model that is used to represent the UI in an abstract fashion. Different types of models have been used in different systems including task models, dialogue models, user models, domain models, and application models. All these models represent the UI at a higher level of abstraction than what is possible with a more concrete representation. The UI developer built these models that were transformed either automatically or semi-automatically to generate the final UI. The generation of the interface was often done via an automated tool using very complex processing. Most of these tools included a set of rules for generation of the interface components as well as for specifying interface style. Thus, extending a model-based development for multiple platforms would require adding new rules to the generation tool for each new interface style.

Although model-based tools provide many advantages over other user interface development tools, one of the main drawbacks of these systems was that the automatically generated interfaces were not of very good quality. It was not feasible to produce good quality interfaces for even moderately complex applications from just data and task models. One more limitation of some of the earlier systems was the lack of user control over the process of UI generation. Having more developer-control over the UI development process and having more usable models could rectify some of the deficiencies of the model-based approaches.

A transformation-based approach allows the developer to have more control over intermediate steps than the model-based approach ever allowed. Furthermore, the transformations done in our approach encompass simpler processing, thus making it easier for the developer to understand the generation process, and potentially even extend the transformation process by creating new transformations.

## 3.4 Markup Languages and the World Wide Web

Since the advent of the World Wide Web in the mid-90s and the emergence of eXtensible Markup Language (XML) as a standard meta-language, a number of different markup languages have emerged for creating UIs for different devices. The foremost among these are HTML [28] for desktop machines, WML [26] for small hand-held devices, and VoiceXML [24] for voice-enabled devices. XML itself is a meta-language that allows the definition of other languages. There have been various other standards developed in conjunction with XML including XSLT that allows XML to be converted to other formats based on some rules.

HTML can be considered a language for multi-platform development, but it only supports platforms that are in the same family i.e., primarily desktop computers. Some efforts have tried to make HTML files available in other devices (e.g. see Transcoding above) but in general HTML has remained tied to desktop computers. The Wireless Markup Language (WML), a part of the Wireless Application Protocol (WAP), is an XML-based language designed primarily for devices with small screen-sizes and limited bandwidth, including cellular phones and pagers. WML uses the metaphor of a deck of cards to represent a UI with a user having to navigate between different cards that are grouped together like a deck. VoiceXML is a markup language for specifying interactive voice response applications. It is designed for creating audio dialogs that feature synthesized speech, digitized audio, recognition of spoken and Dual-Tone Multi Frequency key input, recording of spoken input, telephony, and mixed-initiative conversations. Xforms [27] is the next generation of HTML forms, which intends to greatly enhance the capability of the current forms available in HTML.

While all of these markup languages have almost entirely removed the need to know toolkit, hardware, and operating system specifics for UI development, they have not made a significant contribution towards multi-platform development. The use of these languages for multi-platform development still depends on the client browser implementing the language appropriately. Today, for example, not even basic HTML can be considered multi-platform as the two popular browsers, Microsoft's Internet Explorer and Netscape's Communicator, implement slightly incompatible versions of HTML.

## 4.  UIML

UIML [1, 15, 16] is a declarative XML-based language that can be used to define user interfaces. One of the original design goals of UIML is to "reduce the time to develop user interfaces for multiple device families" [2]. A related design rationale behind UIML is to "allow a family of interfaces to be created in which the common features are factored out" [1]. One of the primary design goals of UIML is to provide a canonical format for describing interfaces that map to multiple devices. In this section we present some of UIML's language features.

```
<?xml version="1.0" ?>
<!DOCTYPE uiml PUBLIC "-//Harmonia//DTD UIML 2.0 Draft//EN"
   "UIML2_0g.dtd">
<uiml>
  <head>...</head>
  <interface>
     <structure>...</structure>
     <content>...</content>
     <behavior>...</behavior>
     <style>...</style>
  </interface>
  <peers>...</peers>
  <template>...</template>
</uiml>
```

Figure 1: Skeleton of a UIML document with the Interface section highlighted

## 4.1  Language Overview

Since the language is XML-based, the different components of a user interface are represented through a set of tags. The language itself does not contain any platform-specific or metaphor-dependent tags. For example, there is no tag like <window> that is directly linked to the desktop metaphor of interaction. UIML uses about thirty generic tags instead. Platform-specific renderers have to be built in order to render the interface defined in UIML for that particular platform. Associated with each platform-specific renderer is a vocabulary of the language widget-set or tags that are used to define the interface in the target platform.

A skeleton UIML document is represented in Figure 1. At the highest level, a UIML document comprises of four components: <head>, <interface>, <peers> and <template>. The <interface> is the only component that is relevant for this discussion (it is shown in bold in Figure 1); information on the others can be found elsewhere [15]. The <interface> element  is the heart of the UIML document in terms of representing the actual user interface. All the UIML elements that describe the UI are present within this element. There may be multiple <interface> elements. The four main components are the following:

a. <structure>: The physical organization of the interface, including the relationships between the various UI elements within the interface, is represented using this element. Each <structure> is comprised of different <part>s. Each part represents the actual platform-specific UI Element and is associated with a single class of UI elements. The term "class" in UIML represents a particular category of UI elements. Different parts may be nested to represent a hierarchical relationship. There might be more than one structure in a UIML document representing different organizations of the same UI.

b. <style>: The style contains a list of properties and values used to render the interface. The properties are usually associated with individual parts within the UIML document through the part-names. Properties can also be associated with particular classes of parts. Typical properties associated with parts for Graphical User Interfaces (GUIs) could be the background color, foreground color, font, etc. It is also possible to have multiple styles within a single UIML document to be possibly associated with multiple structures or even the same structure. This facilitates the use of different styles for different contexts

c. <content>: This represents the actual content associated with the various parts of the interface. A clean separation of the content from the structure is useful when different content is needed under different contexts. This feature of UIML is very helpful when creating interfaces that might be used in multiple languages. An example of this is a UI in French and English, for which separate content is needed.

d. <behavior>: The behavior of an interface is specified by enumerating a set of conditions and associated actions within rules. UIML permits two types of conditions. The first condition is when an event occurs, while the second is true when an event occurs and the value of some data associated with the event is equal to a certain value. There are four kinds of actions that occur. The first action is to assign a value to a part's property. The second action is to call an external function or method. The third is to fire an event and the fourth action is to restructure the interface.

A detailed discussion about the language features can be found in the UIML language specification [15], available at http://www.uiml.org. A discussion of the language design issues can be found in Phanouriou's dissertation [16].

Currently, there are platform-specific renderers available for UIML for a number of different platforms. These include Java, HTML, WML, and VoiceXML. Each of these renderers has a platform-specific vocabulary associated with it to describe the UI elements, their behavior and layout. The formal definition of each vocabulary is available at http://www.uiml.org/toolkits. The table below shows some of the vocabularies publicly available at the time of publication.

Table 1. UIML Vocabularies available from http://www.uiml.org/toolkits as of August 2001

| Generic Vocabulary Available for Platform |
| --- |
| W3C's Cascading Style Sheets (CSS) |
| W3C's Hypertext Markup Language (HTML) v4.01 with the frameset DTD and CSS Level 1 |
| Java™ 2 SDK (J2SE) v1.3, specifying AWT and Swing toolkits |
| A single, generic (or multi-platform) vocabulary for creating Java *and* HTML user interfaces |
| VoiceXML Forum's VoiceXML v1.0 |
| WAP Forum's Wireless Markup Language (WML) v1.3 |

The mechanism that is currently employed for creating UIs with UIML is that the UI developer uses the platform-specific vocabulary to create a UIML document that is rendered for the target platform. These renderers can be downloaded from http://www.harmonia.com.

The platform-specific vocabulary for Java uses AWT and Swing class names as UIML part names. The platform-specific vocabulary for HTML, WML, and VoiceXML uses HTML, WML, and VoiceXML tags as UIML part names, respectively. This enables the UIML author to create a UI in UIML that is equivalent to a UI that is possible in Java, HTML, WML, and VoiceXML. However, the platform-specific vocabularies are not suitable for a UI author that wants to create UIML documents that map to *multiple* target platforms. For this a *generic* vocabulary is needed. To date, one generic vocabulary has been defined, *GenericJH*, which maps to both Java Swing and HTML 4.0. The next section describes how a generic vocabulary is used with UIML.

## 5.  GENERIC VOCABULARY AND TRANSFORMATIONS

From Section 2 above, we recall that the definition of *family* refers to multiple *platforms* that share common layout capabilities. Different *platforms* within a *family* often differ on the *toolkit* used to build the interface. Consider for example, a Windows OS machine capable of displaying HTML using some browser and capable also of running Java applications. Both of these use different *toolkits*, thus making it impossible to write an application in one and have it execute in the other, even though they both run on the same hardware device using the same operating system. For these particular cases, we have built into UIML support for *generic vocabularies*.

A *generic vocabulary* of UI elements, used in conjunction with UIML, can describe any UI for any *platform* within its *family*. The vocabulary has two objectives: first, to be powerful enough to accommodate a *family* of devices, and second, to be generic enough to be used without having expertise in all the various *platforms* and *toolkits* within the *family*.

We have identified and defined a set of generic UI elements including their properties and events for desktop applications. Ali, *et al.* [3] provides a more detailed description of this *generic vocabulary*. The table below shows some of the part classes in the *generic vocabulary* for the *family* including HTML 4 and Java Swing.

Table 2. Example of a Generic Vocabulary

| Generic Part | UIML Class Name | Generic Part | UIML Class Name |
| --- | --- | --- | --- |
| Generic top container | G:TopContainer | Generic Label | G:Label |
| Generic area | G:Area | Generic Button | G:Button |
| Generic Internal | G:InternalFrame | Generic Icon | G:Icon |
| Generic Menu Item | G:Menu | Generic Radio Button | G:RadioButton |
| Generic Menubar | G:MenuBar | Generic File Chooser | G:FileChooser |

## 5.1  Transformations

A *generic vocabulary* is useless without a corresponding set of transformations that convert the user interface from the generic terms to a particular *toolkit*. These transformations are basically a conversion from generic UIML to platform-specific UIML. Both of these representations be represented as trees since they are XML-based. The platform specific UIML is then *rendered* using an existing UIML *renderer*. There are several types of transformations that are performed:

- Map a generic class name (e.g., G:TopContainer) to one or more parts in the target platform. For example, a G:TopContainer is mapped to the following sequence of parts in HTML:

```
<html>
  <head>...
    <title>...
    <base>...
    <style>...
    <link>...
    <meta>...
  <body>...
```

  In contrast, G:TopContainer is mapped simply to one part, JFrame, in Java.

- Map properties of the generic part to the proper platform-specific part.

- Map events for the generic part to the proper platform-specific part, and translate rules in the <behavior> element of UIML to use the platform-specific events.

In order to allow a UI designer to fine tune the UI to a particular platform, the generic vocabulary contains platform-specific properties. This are used when one platform has a property that has no equivalent on another platform. In the generic vocabulary, these property names are prefixed by J: or H: for mapping to Java or HTML only. The transform engine automatically identifies which target part to associate the property with, in the event that a generic part (e.g., G:TopContainer) maps to several parts (e.g., 7 parts for HTML). This is also done for events that are specific to one platform. In this way, the generic vocabulary is not a lowest-common-denominator approach. The generic UIML file would then contain three <style> elements. One is for cross-platform style, one for HTML, and one for Java UIs:

```
<uiml>
...
  <style id="allPlatforms">
    <property id="g:title">My User Interface</property>
  </style>
  <style id="onlyHTML" source="allPlatforms">
    <property id="h:link-color">red</property>
  </style>
  <style id="onlyJava" source="allPlatforms">
    <property id="j:resizable">red</property>
  </style>
...
</uiml>
```

In the example above, both a web browser and a Java frame have a title, which is *My User Interface*. However, only web browsers can have the color of their links set, so property h:link-color is used only for HTML UIs. Similarly, only Java UIs can make themselves non-resizable, so the j:resizable property applies only to Java UIs.

When the UI is rendered, the renderer will choose exactly one <style> element. For example, an HTML UI would use *onlyHTML*. This <style> element specifies in its *source* attribute the name of the common, shared *allPlatforms* style, so the *allPlatforms* style is shared by both the HTML and Java style elements.

## 6. TRANSFORMATION-BASED UI DEVELOPMENT

A transformation-based UI development places the developer in unfamiliar territory. Developers are accustomed to have total control over the language and the specification of the user interface elements. A transformation-based environment asks the developer to provide a high-level description of the interface and to "trust" the result. This is one of the common limitations of code-generators and model-based UI systems. TIDE was developed at Virginia Tech and is freely available at Harmonia's website.

To address this limitation, we have developed a Transformation-based Integrated Development Environment (TIDE) for UIML. In TIDE, the developer writes UIML code and the IDE generates the interface, as expected. However, the relationship between UIML code and its resulting interface component is explicitly shown. This section briefly shows how TIDE works.

The TIDE application was built on the idea that developers creating an interface in an abstract language, such as UIML, which will be translated into one or more specific languages, undergo a process of trial and error. The developer builds what he or she thinks will be appropriate in UIML, renders to the desired language(s), and then makes changes as appropriate. As an environment designed to help support this process, this application shows the developer three things: the original, abstract interface (in source code form), the resultant interface after rendering, and the relationship between elements in the two. Figure 2 below shows two pictures of the TIDE environment.

The application uses Harmonia's LiquidUI product suite, version 1.0c. Harmonia's product was not changed in any way, and was used by the application to render from the original UIML to Java. The developer may open and close files, view the original UIML source code as plain text or as a tree, and make changes from the tree view. The developer can also re-render at any time (by pressing the red arrow in the center of the window).

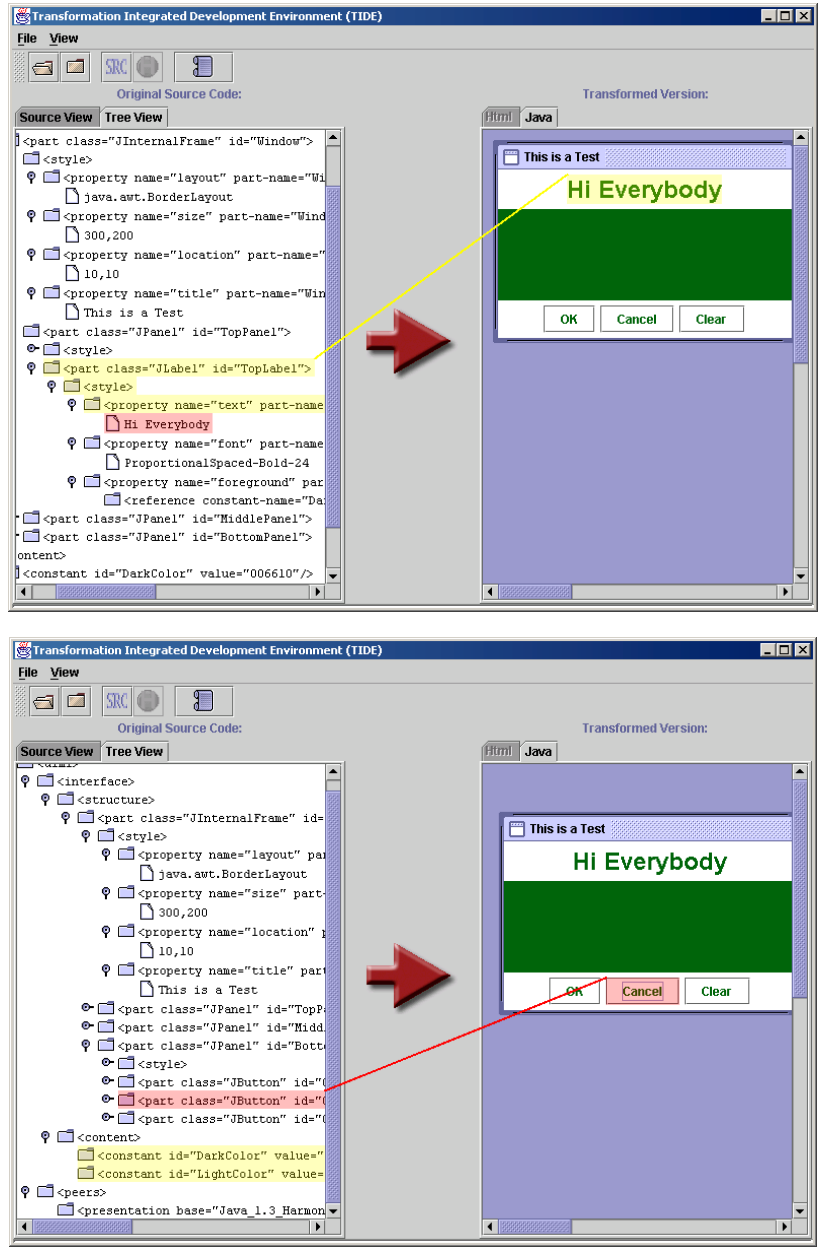

Figure 2. Selection in UIML Code

The relationship between UIML code and the generated interface is made explicit as shown in Figure 2 above. The developer may click on a node in the UIML tree view and the corresponding element on the graphical user interface is highlighted on the right side. The same is true if the developer clicks on a component of the graphical user interface. In the right hand side of Figure 2, the developer has clicked on the cancel button (middle of the three buttons) and the corresponding code is highlighted on the UIML tree-view.

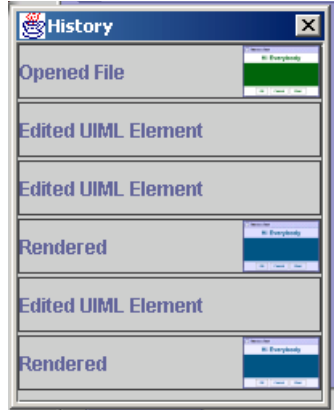

Figure 3. History Window in TIDE

TIDE makes very easy to explore the different UIML elements and the effect they have on the generation of the interface. A UIML element's property (e.g. the color of a button) can be edited in place within the tree view. TIDE even supports a history window that keeps track of different changes made to the interface. Each line in the history window (see Figure 3 above) shows a small screen image of the interface at that point in the development cycle. This allows the developer to quickly switch between alternative versions of the interface, thus encouraging more exploration of UIML.

# 7. SAMPLE APPLICATIONS

This section shows two sample UIML applications. The first one is intended to show some of the parts of UIML at work. For economy of space, we have not included the <style> and <behavior> sections of the UIML document. See Ali [3] for other UIML examples.

```
<?xml version="1.0"?>
<!DOCTYPE uiml PUBLIC "-//Harmonia//DTD UIML 2.0 Draft//EN"
  "UIML2_0g.dtd">
<uiml>
 <head>
  <meta name="Purpose" content="Data Collection Form"/>
  <meta name="Author"  content="Farooq Ali"/>
 </head>
<interface name="DataCollectionForm">
 <structure>
  <part name="RequestWindow" class="G:TopContainer">
    <part name="EBlock1" class="G:Area">
     <part name="TitleLabel" class="G:Label"/>
     <part name="FirstName" class="G:Label"/>
     <part name="FirstNameField" class="G:Text"/>
     <part name="LastName" class="G:Label"/>
     <part name="LastNameField" class="G:Text"/>
     <part name="StreetAddress" class="G:Label"/>
     <part name="StreetAddressField" class="G:Text"/>
     <part name="City" class="G:Label"/>
     <part name="CityField" class="G:Text"/>
     <part name="State" class="G:Label"/>
     <part name="StateChoice" class="G:List"/>
     <part name="Zip" class="G:Label"/>
     <part name="ZipField" class="G:Text"/>

     <part name="OKBtn" class="G:Button"/>
     <part name="CancelBtn" class="G:Button"/>
     <part name="ResetBtn" class="G:Button"/>
    </part>
   </part>
  </structure>
```

We can see from the snippet of code above that the *generic vocabulary* comprises of UI elements with names that are applicable to multiple *platforms* within this *family*. The code shown above represents multiple *platforms* (Java and HTML) for a single *family*.

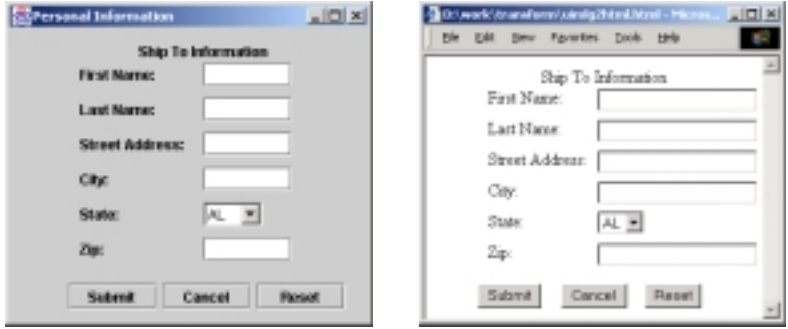

Figure 4: Screen-shots of a sample form in Java Swing (left) and HTML (right).

The second example we show is a more complex UI. For this, we show only the generated interfaces for economy of space. The image at the top is the UIML rendered for Java Swing. The one at the bottom is rendered for Internet Explorer v5.5.

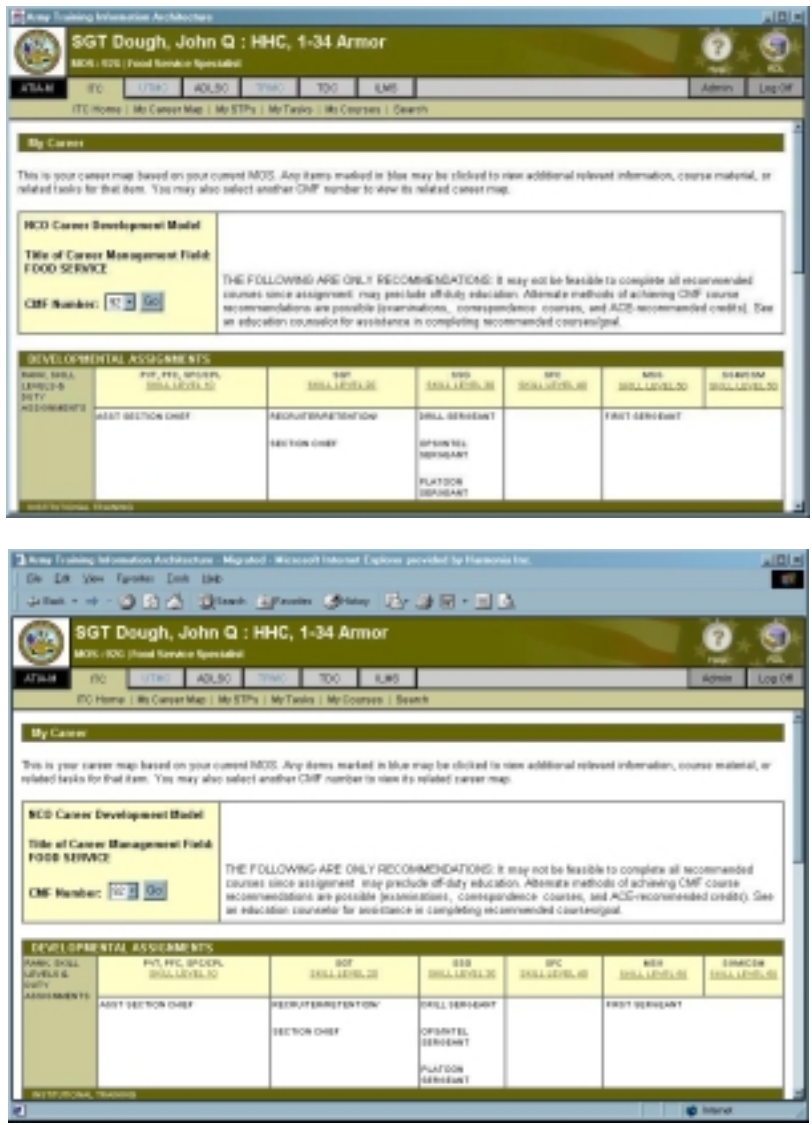

Figure 5: Screen-shots of a complex application for Java and HTML built with UIML

# 8.   FRAMEWORK FOR MULTI-PLATFORM UI

The UIML language and tools presented above are not sufficient to properly address the problem of interface development for multiple platforms. In some cases, the differences between display layouts were found to be too significant to simply create one UIML file for one particular platform and expect it to be rendered on a different platform. Therefore, we are adding a more abstract layer to support development of interfaces for a wide variety of platforms. This section describes preliminary work on how this higher-abstraction layer would work within UIML. The approach taken is to have a higher level UIML representation with its own (and new) generic vocabulary.  We are using the traditional usability engineering process to help us build a new methodology that could be used to build interfaces for multiple-platforms.

The concept of building multi-platform UIs is relatively new. To envision the development process, we followed an existing approach from the usability engineering (UE) literature. One such approach [9] identifies three different phases in the UI development process: interaction design, interaction software design and interaction software implementation. Interaction design is the phase of the usability engineering cycle in which the "look and feel" and behavior of an UI is designed in response to what a user hears, sees or does. In current Usability Engineering practices, this phase is highly platform-specific.

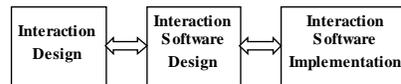

Figure 6: Traditional Usability Engineering Process for one platform

Once the interaction design is done, the interaction software is designed, which involves making decisions about the UI toolkit(s), widgets, positioning of widgets, colors, etc. Once this design is done, the software is implemented Unfortunately, this traditional view of interaction design is highly platform-specific and works well for a single platform. The stage of interaction design has to be further split into two distinct phases for multiple platforms: a platform-independent interaction design and platform-dependent interaction design. This phase will lead to different platform-specific interaction software designs, which in turn lead to platform-specific UIs. This process is illustrated in Figure 7.

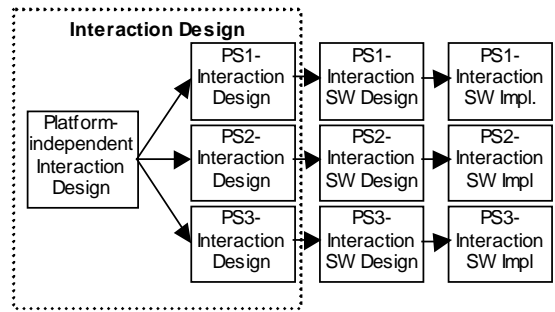

Figure 7: Usability Engineering process for multiple platforms

## 8.1   UIML-Based Usability Engineering Process

We have developed a framework that is very closely related to the UE process discussed above. The main building blocks within this framework are the *Logical Model, Family Model* and the *Platform-Specific Model*. The *logical model* is the new model that we are creating, and it is briefly described below. The *family* model is a model describing interfaces for a specific *family* using its corresponding *generic vocabulary*. The *Platform-Specific Model* describes a user interfaces using the vocabulary associated with a particular toolkit. There is also a new process of transformations needed that transforms the *logical model* to a particular *family model*. Thus, we have one new level of representation, the *logical model*, and one new process of transformations. The resulting diagram is shown in Figure 8.

The main objective of the *logical model* is to capture the UI description at a higher level of abstraction than is possible by any family-specific model. The logical model is a collection of logical constructs that represent the UI in some abstract fashion. We are considering using a hybrid task/domain model that can be transformed to the physical model. Research on this representation is underway.

The transformation from the *logical model* to a particular *family model* has to be developer-guided and cannot be fully automated. By allowing the UI developer to intervene in the transformation and mapping process, it is possible to ensure usability. One of the main problems of some of the earlier model-based systems was that a large part of the UI generation process from the abstract models was fully automated, removing user-control of the process. Another way of describing this problem is the "mapping problem" as described by Puerta [18]. We want to eliminate this problem by having the user guide the mapping process. Once the user had identified the mappings, the system will generate a *family model* based on the target family and the user mappings. Research is underway on these transformation algorithms and the process to specify the mappings. We are integrating the transformation specification within TIDE, thus having a single integrated environment for the developer; our goal is to avoid the feelings of loss of control that plagued earlier model-based approaches.

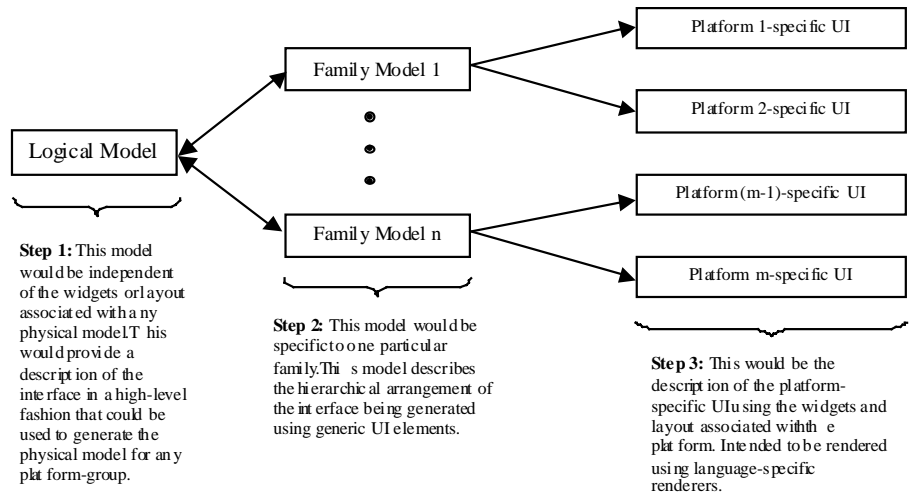

Figure 8: The overall framework for building multi-platform UIs using UIML

# 9. CONCLUSIONS

In this paper we have shown some of our research in UIML. We have particularly focused on the overall approach we have taken to generate multi-platform user interfaces. We are using a single language, UIML, to provide the multi-platform development support needed. This language is extended via the use of alternate vocabularies and transformation algorithms. Our approach allows the developer to build a single specification for a *family* of devices. UIML and its associated tools transform this single representation to multiple platform-specific representations that can then be rendered in each device. We have presented our current research in extending UIML to allow building of interfaces for very-different platforms, such as VoiceXML, PDAs and desktop computers.